\RequirePackage{fix-cm}
\documentclass[aps,pra,reprint,superscriptaddress]{revtex4-2}

\RequirePackage{graphicx}
\usepackage{graphicx}% Include figure files
\usepackage{dcolumn}% Align table columns on decimal point
\usepackage{bm}% bold math
\usepackage{lipsum}
\usepackage{subfigure}

\usepackage{tikz}
\usepackage{ctable}
\usetikzlibrary{arrows.meta}
\usepackage{pgfplots}
\usepgfplotslibrary{colormaps}
\pgfplotsset{compat=1.8}
\usetikzlibrary{pgfplots.groupplots,pgfplots.fillbetween}

\usepackage[colorlinks=true,linkcolor=black,citecolor=blue,urlcolor=blue,]{hyperref}
\def \esym{$E_{\rm sym}(\rho)$}

\begin{document}

\title{The trigger system for the CSR external-target experiment}
\author{Dong Guo} % guodong19@mails.tsinghua.edu.cn
\email[]{guodong19@mails.tsinghua.edu.cn}
\affiliation{Department of Physics, Tsinghua University, Beijing 100084, China}

\author{Haoqian Xyu} % xhqlxx@mail.ustc.edu.cn
\affiliation{Department of Modern Physics, University of Science and Technology of China, Hefei 230026, China}

\author{DongDong Qi} % qidd@mail.ustc.edu.cn
\affiliation{Department of Modern Physics, University of Science and Technology of China, Hefei 230026, China}

\author{HeXiang Wang} % wanghexiang@mail.ustc.edu.cn
\affiliation{Department of Modern Physics, University of Science and Technology of China, Hefei 230026, China}

\author{Lei Zhang} % zl1998@mail.ustc.edu.cn
\affiliation{Department of Modern Physics, University of Science and Technology of China, Hefei 230026, China}

\author{Zhengyang Sun} % zgkdszy@mail.ustc.edu.cn
\affiliation{Department of Modern Physics, University of Science and Technology of China, Hefei 230026, China}

\author{Zhi Qin} % qinz18@mails.tsinghua.edu.cn
\affiliation{Department of Physics, Tsinghua University, Beijing 100084, China}

\author{Botan Wang} % wbt19@mails.tsinghua.edu.cn
\affiliation{Department of Engineering Physics, Tsinghua University, Beijing 100084, China}

\author{Yingjie Zhou} % zhouyingjie@mail.ustc.edu.cn
\affiliation{Department of Modern Physics, University of Science and Technology of China, Hefei 230026, China}

\author{Zekun Wang} % wangzekun@impcas.ac.cn
\affiliation{Institute of Modern Physics, Chinese Academy of Science, Lanzhou 730000, China}

\author{Yuansheng Yang} % yangyuansheng@impcas.ac.cn
\affiliation{Institute of Modern Physics, Chinese Academy of Science, Lanzhou 730000, China}

\author{Yuhao Qin} % qinyh18@mails.tsinghua.edu.cn
\affiliation{Department of Physics, Tsinghua University, Beijing 100084, China}

\author{Xianglun Wei} % weixl@impcas.ac.cn
\affiliation{Institute of Modern Physics, Chinese Academy of Science, Lanzhou 730000, China}

\author{Herun Yang} % yanghr@impcas.ac.cn
\affiliation{Institute of Modern Physics, Chinese Academy of Science, Lanzhou 730000, China}

\author{Yuhong Yu} % yuyuhong@impcas.ac.cn
\affiliation{Institute of Modern Physics, Chinese Academy of Science, Lanzhou 730000, China}

\author{Lei Zhao} % zlei@ustc.edu.cn
\affiliation{Department of Modern Physics, University of Science and Technology of China, Hefei 230026, China}

\author{Zhigang Xiao} % xiaozg@mail.tsinghua.edu.cn
\email[]{xiaozg@mail.tsinghua.edu.cn}
\affiliation{Department of Physics, Tsinghua University, Beijing 100084, China}

\begin{abstract} 
A trigger system has been designed and implemented for the HIRFL-CSR external target experiment (CEE), the spectrometer for studying nuclear matter properties with heavy ion collisions in the GeV energy region. The system adopts master-slave structure and serial data transmission mode using optical fiber to deal with different types of detectors and long-distance signal transmission. The trigger logic can be accessed based on command register and controlled by a remote computer. The overall field programmable gate array (FPGA) logic can be flexibly reconfigured online to  match the physical requirements of the experiment. The trigger system has been tested in beam experiment. It is demonstrated that the trigger system functions correctly and meets the physical requirements of CEE. 

%Use showkeys class option if keyword

\end{abstract}

\maketitle

%%%%%%%%%%%%%%%%%%%%%%%%%%%%%%%%%%%%%%%%%%%%%%%%%%%%%%%%%%%%%%%%%%%

\section{Introduction}\label{sec. I}

The QCD phase structure at high net baryon density is drawing increasing attention in the field of high energy heavy ion physics \cite{Luo2022PropertiesBook}. Despite of enormous progress in studying the QCD phase diagram ~\cite{Braun2007QGP}, the enriched structure on the diagram is still an open question drawing wide attention from the community of nuclear physics. Recently, the STAR experiment collected data  in the energy range from 200 to 7.7 GeV ~\cite{Luo2015Energy, Luo2016Exploring, Abdallah2021Cumulants}, the cumulants of net proton and proton $\kappa\sigma^{2}$ (or $C_{4}/C_{2}$) as a function of  $\sqrt{s_{NN}}$ were investigated in central Au+Au collisions, and a totally different behavior of $\kappa\sigma^{2}$  at lower collision energies has been observed ~\cite{Abdallah2021Cumulants, STAR2021Nonmonotonic}. It indicates the opportunities  for accurate measurements of observables at high net baryon density region. Particularly in hadron phase, the determination of the isovector sector of the EoS of asymmetric nuclear matter, i.e., the density dependent nuclear symmetry energy, \esym, receives increasing interest too, because this not-well-constrained quantity plays an important role in radioactive nuclear structure, isospin dynamics in heavy ion reactions, liquid-gas phase transition in asymmetric nuclear matter and neutron star structure etc ~\cite{BaoAn2013Constraining, 2017BaoNuclear, Abbott2018GW170817, 2021BaoProgress}. The present theoretical and experimental work shows that the symmetry energy is still uncertain in the supersaturated density region ~\cite{Xiao2009Circumstantial, Xiao2014Probing, Russotto2011Symmetry, Xie2013Symmetry, Yangyang2021Insights}. In order to constrain stringently \esym~ particularly at  high densities, great efforts are taken in worldwide laboratories.

The CEE is a generally purposed spectrometer under construction on the Cooling Storage Ring at the Heavy Ion Research Facility in Lanzhou (HIRFL-CSR), China. It aims at the studies of the structure of QCD phase diagram at high net-baryon density region ~\cite{LiMing2016Conceptual, guo2023studies}. The physical programs in plan include the search for the signals unraveling the existence of the critical end point (CEP) of QCD phase transition, the constraint of \esym~ at  supra-saturation density region using various observables, and the studies of the properties of hypernuclei  as well as some other  issues at the  frontiers of nuclear physics ~\cite{Yuan2020Present}. It is a fixed target experiment and covers more than $2\pi$ space in center of mass system.  The highest beam energy achievable is 0.5 and 2.8 GeV/u  for uranium and proton beam, respectively. CEE consists of various subsystems including tracking detectors, time of flight (TOF) detectors, zero-degree counters {\it etc}. It contains  a total number of channels of about 20 K, running  with a maximum event rate of 10 kHz.    

The trigger system, playing significant role in large-scale experiments like CEE in nuclear physics, sees rapid developments in recent years due to the increasing application of  the field programmable logic gate array (FPGA) technology ~\cite{Monmasson2007FPGA, Pozniak_2010FPGAbased}. FPGA technology has obvious advantages of compact space, low power consumption, strong adaptability and good scalability ~\cite{Bagliesi2010TOTEM}, as well as of  convenient operation for remote control ~\cite{Volker2004trigger, Bagliesi2010TOTEM, LiMin2016trigger}. Meanwhile, functions including digital signal processing algorithms ~\cite{Szadkowski2006algorithm, Imbergamo2008Nuclear, Liuyinyu2018Implementation}, time digitization design ~\cite{Mircea2005Nuclear, Suwada2015Wide, Lu2021Readout}, track reconstruction of charged particles ~\cite{Bartz_2020, Dong2022FPGA}, can be implemented in the FPGA-based  trigger system, as demonstrated by the wide applications in ATLAS ~\cite{Garvey2003FPGA, Andrei2006FPGA}, ALICE ~\cite{Rinella2007ALICE, Engel2017FPGA}, and CMS ~\cite{Jeitler2013Nuclear, Cela-Ruiz_2017} etc.  

Adapting the up-to-date FPGA technologies, the trigger system of CEE has been designed to fulfill  the following functions: 

1) In the beam experiment, it is able to provide the experimental trigger signals, selecting the physical collision events on the target and suppressing  the background events off-target. 

2) In the debugging process of each detector, it can provide laser testing and cosmic ray trigger signals. 

3) It carries out the task of sending and receiving the signals and commands of the global clock synchronization. 

4) It has scalability and reloadability to realize data flow and command interaction with the DAQ system. 

In this  paper, we introduce the design of trigger system in CEE. Section \ref{sec. II} introduces the physical requirements and logic of the trigger system, section \ref{sec. III} introduces the design of the trigger system, section \ref{sec. IV} introduces the beam test results, and section \ref{sec. V} is the summary.

\section{Trigger conditions for CEE} \label{sec. II}

\subsection{Overall design of CEE } \label{sec.II.1} 

\begin{figure}[htb]
\centering
% \hspace{-0.7cm}
\includegraphics[width=0.40\textwidth]{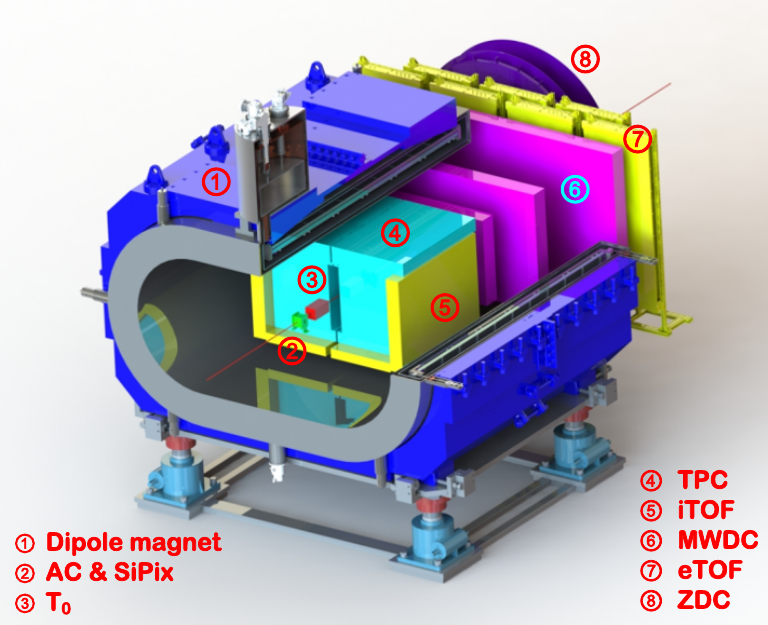}
\caption{(Color online) The conceptual design of the CEE.}
\label{CEE}
\end{figure}

The main component of CEE is a large-gap dipole magnet, housing the tracking detectors in the magnetic field. The time projection chamber (TPC) with the sensitive volume of ${\rm 120(x)\times 80(y)\times 90(z) cm^3}$ ~\cite{Li2016Simulation}, read out by gaseous electron multiplication (GEM) and induction pad plane,    is installed in the center of the magnetic field. The TPC consists of two independent sets of detectors, which are symmetrically distributed along the beam direction, leaving a blank area for the beam to pass through. The inner time-of-flight (iTOF) detector ~\cite{Wang2022CEE} consists of multi resistive plate counters (MRPCs) covering the left, right and bottom sides of TPC. Three multi wire drift chamber (MWDC) tracking detectors  of different sizes are installed covering the forward angle of $5^\circ<\theta_{\rm lab}<30^\circ$ on the downstream side of the TPC. Two sets of MWDC are placed inside the atmospheric gap magnet, and one set of MWDC is placed outside the magnet. An inactive area is made to allow  the beam passing  through the MWDC array. The end cap time of flight detector (eTOF) ~\cite{Botan2020eTOF} is located behind the MWDC, providing the arrival timing signal for the light charged particles in forward rapidity region. The starting time signal is provided by  ${\rm T_0}$  installed on the upstream of the target. The high performance time digitization modules (TDMs) are adopted  to obtain high precision timing information ~\cite{Lu2021Readout}. The zero angle calorimeter (ZDC) ~\cite{Zhu2021Prototype} is located behind all other detector components and the magnetic field, in order to deliver the information related to the reaction plane and the centrality by measuring  the charged-resolved spectators in very forward region.  An active collimator  detector (AC) and a silicon pixel  detector (SiPix) ~\cite{Liu2023Design} are installed between the target and ${\rm T_0}$ on the beam line to suppress the background events induced by the projectile with upstream materials and to record the vertex of the beam, respectively. The overall design of CEE is shown in FIG.\ref{CEE} for the details. 

%% Fig.CEE Structure

\subsection{The physical requirements of trigger event} \label{sec.II.1}

\begin{figure}[htb]
\centering
% \hspace{-0.7cm}
\includegraphics[width=0.48\textwidth]{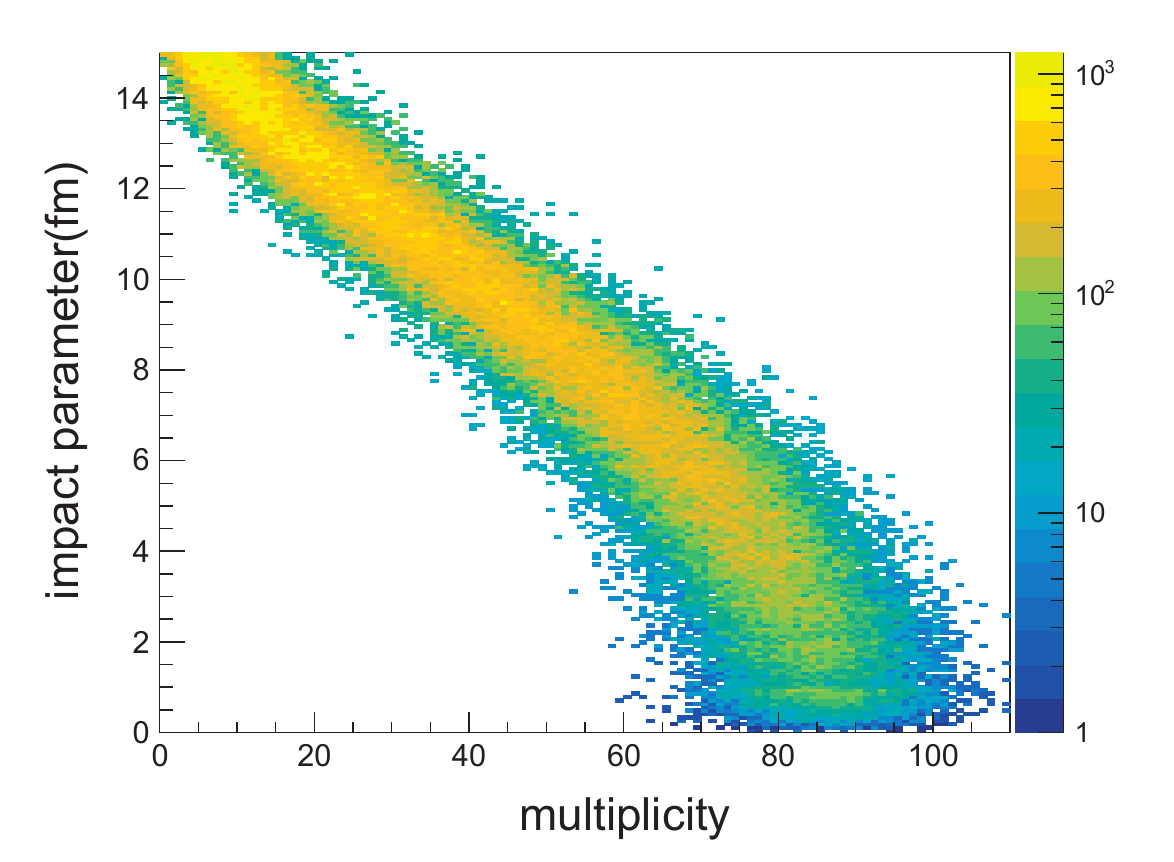}
\caption{(Color online) The scattering plot of the multiplicity recorded the TOF MRPCs {\it vs.} the impact parameter.  }
\label{TOF_M-b}
\end{figure}

For the heavy ion collisions  in the GeV/u energy region, the rough selection of the event geometry (impact parameter)  is the main goal of the trigger system in beam experiment. In general, the trigger conditions are set by the multiplicity of charged particles. The high (low) multiplicity corresponds to the central (peripheral) collisions. Even though the fast trigger detectors covers part of the phase space, the partial multiplicity correlates to the total multiplicity and hence measures the event centrality. Using JET AA Microscopic Transport Model (JAM) ~\cite{Nara1999Relativistic} as the event generator in the framework of CEEROOT developed for experimental simulations and data analysis, one can calculate the multiplicity in iTOF and eTOF as a function of impact parameter $b$, as shown in  Fig.  \ref{TOF_M-b}. It is clearly seen that  the multiplicity of charged particles covered by iTOF and eTOF  has a good reverse correlation with the impact parameter, indicating that the combination of  iTOF and eTOF can give a rough selection of the impact parameter. Pattern recognition of the multiplicity distribution on eTOF can further delivers the centrality with better precision ~\cite{wang2023determination}, this ability can be developed in offline analysis.

%% Fig.TOF_M-b

\subsection{Trigger logic setting} \label{sec.II.2}

Using the multiplicity as an observable characterizing the event geometry, the triggers signal in beam experiments is provided by AC, ${\rm T_0}$, iTOF and eTOF, as described below. The signal firing ${\rm T_0}$ indicates the beam pass through. The multiplicity over a certain threshold recorded on iTOF and eTOF detector in coincidence then characterizes the reactions on the beam path. In order to suppress the off-target reaction events on the upstream side, the AC is used as a veto in the trigger scheme. Hence, the CEE global trigger condition for beam experiment can be written in

\begin{equation}
    \rm GTRG = T_0 \times iTOF \times eTOF \times \bar{AC}
\end{equation}

In more detail, the correspondence between the reaction events and the firing matrix in the four detectors are listed in Tab.\ref{Trigger event selection}.

% Table 1.Trigger event selection
\begin{table*}[!hbtp] 
\caption{(Color online) The detector firing response matrix for various  physical events, ``$\surd$" (``$-$") indicates that the detector responses (not) to the corresponding event type.}
\label{Trigger event selection}
\setlength\tabcolsep{13.5pt}
\centering
\footnotesize
\begin{tabular}{ccccc}
\toprule
Event type & $T_{0}$ & eTOF & iTOF & AC \\[1.0pt]
\midrule
Very peripheral reaction & $\surd$ & $\surd$ & $-$ & $-$  \\[1.0pt]
Off-target events upstream & $\surd$ & $\surd$ & $\surd$ & $\surd$ \\[1.0pt]
Off-target events after TPC & $\surd$ & $\surd$ & $-$ & $-$\\[1.0pt]
Events on target& $\surd$ & $\surd$ & $\surd$ & $-$  \\[1.0pt]
\bottomrule
\end{tabular}
\centering
\footnotesize
\end{table*}

Further, in order to classify different  types of physical events characterized by the impact parameter,  three multiplicity thresholds are introduced  in the trigger selection logic, including  the high noise threshold $M_{\rm h}$, the event classification threshold $M_{\rm e}$ and the low noise threshold $M_{\rm l}$. If the multiplicity $M_{\rm TOF}$ recorded in iTOF and eTOF is lower than  $M_{\rm l}$ or unphysically higher than $M_{\rm h}$, the events are regarded as noise events which are not of interest and will not be triggered. When the multiplicity  $M_{\rm TOF}$ satisfied $M_{\rm l}<M_{\rm TOF}<M_{\rm e}$, it represents a minimum bias  event, whereas the multiplicity  $M_{\rm TOF}$ satisfied $M_{\rm e}<M_{\rm TOF}<M_{\rm h}$, it represents a central or semi-central collision.  

\section{Design of trigger system} \label{sec. III}
\subsection{System architecture and electronics} \label{sec.III.1}

\begin{figure}[htb]
\centering
% \hspace{-0.7cm}
\includegraphics[width=0.48\textwidth]{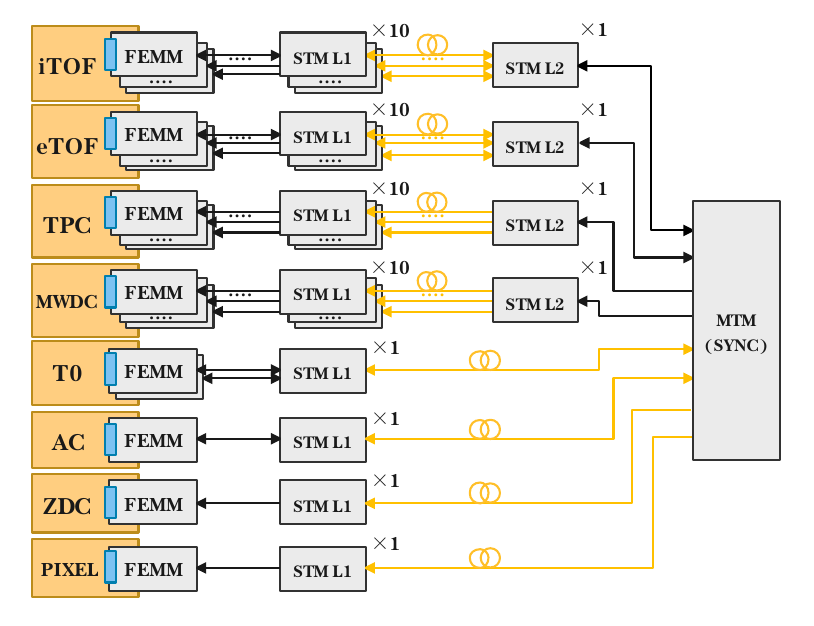}
\caption{(Color online) CEE trigger system master-slave multi-layer structure diagram. The figure shows the number of electronics required for each subsystem in the experiment}
\label{master-slave}
\end{figure}

The CEE trigger system is designed in a master-slave multi-hierarchy structure, maintaining  good  expansion capability. The firing signal uplink from trigger detectors and the trigger signal delivery  to all subsystems are implemented by optical fiber long distance transmission. As shown in  Fig.\ref{master-slave}, the multi-layer architecture of the trigger system  consists of the front-end measurement module (FEMM), the Slave Trigger Module (STM) and the Master Trigger Module (MTM). All trigger electronics of the CEE use a global synchronous  clock with 40 MHz frequency. The iTOF and eTOF  systems adopt a 2-level trigger structure, functioning in the multiplicity calculation and summation. The tracking detector system MWDC and TPC adopt 2-level trigger structure because of the large number of channels. The ${\rm T_0}$, AC, ZDC and SiPix systems adopt a 1-level trigger structure. In the trigger scheme,  ${\rm T_0}$, AC, iTOF and eTOF systems participate the uplink for the trigger signal input, and all systems complete the global trigger signal downlink transmission.

%% Fig.master-slave

The hardware of the MTM mainly includes the electrical connector and optical port for signal transmission, the FPGA chips to implement the event selection algorithm  and the power module. LEMO differential socket is adopted for the 40 MHz clock input electrical port. The interface used for signal transmission with STM L2 of the subsystem adopts LEMO double-layer single-ended socket. A total of 16 single-ended LEMO sockets and 9 Small form pluggable (SFP) can meet various signal transmission requirements of MTM. 

\begin{figure*}[htb]
\centering
% \hspace{-0.7cm}
\includegraphics[width=0.85\textwidth]{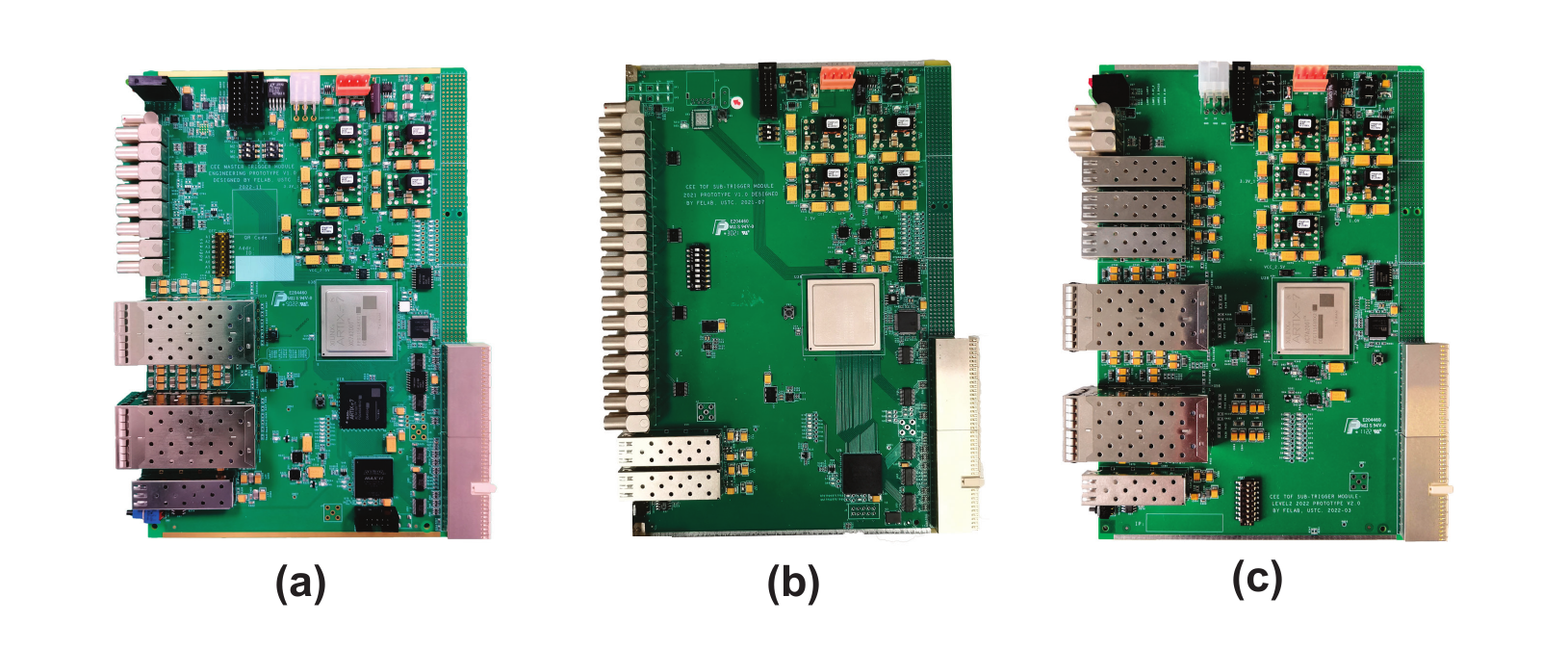}
\caption{(Color online) The real photo of trigger electronics. (a)The real photo of MTM, (b)The real photo of STM L1, (c)The real photo of STM L2}
\label{hardware}
\end{figure*}

The FPGA applied for the MTM is XC7A200TFFG1156 of Xilinx Artix 7 series ~\cite{Artix-7, Xilinx-7}. This type of FPGA integrates 16 GTP and supports serial data transmission rate from 500 Mbps to 6.6 Gbps. The total number of pins is 1156, providing 215360 logical units. The sufficient memory space can meet the needs of event selection algorithm and trigger information cache. The physical photo of the MTM board is shown in FIG.\ref{hardware} (a).

%% Fig.Trigger electronics hardware

The STM mainly completes the information exchange with  FEMM, the signal transmission between the 2-level trigger modules, and the communications with the MTM to receive and send signals. The STM L1 is equipped with 32 single-ended LEMO sockets and 2 SFP, while the STM L2 is designed to have 4 single-ended LEMO sockets and 11 SFP. Xilinx Artix 7 series XC7A200TFFG1156 is selected for both the type of STM FPGA ~\cite{Artix-7, Xilinx-7}. For the TOF subsystem, STM L1  receives  the hit signals  from the TDM  and calculates the local multiplicity to be sent to STM L2 via optical fiber transmission. Then from STM L2 the summation operation to the received multiplicity is done and the result is sent to the MTM to construct the global trigger signal.  The physical photo of STM L1 and STM L2 are shown in FIG.\ref{hardware} (b) and (c).

The trigger system adopts the Gigabit Transceiver with Low Power (GTP) integrated by Xilinx in Artix 7 FPGA to implement the function of optical fiber long-distance signal transmission. In GTP, 8B/10B encoding mode of series-parallel/parallel-series conversion is used to ensure the DC balance in serial data transmission. After the serial signal is generated, the photoelectric conversion is further completed, and the electrical signal is converted into an optical signal sent through the SFP. FTLF8519P2BNL with SFP from Finisar ~\cite{Finisar-SFP} is selected in the experiment, which has a transmitter and a receiver. The maximum rate of 2.125 Gbps serial data transceiver within 500 meters can be achieved. In the logic function, the logic input electrical signal and the output electrical signal of GTP are 16-bit parallel signals, and when the electrical signal is converted into the optical signal in the logic, the serial signal of 1-bit needs to do series-parallel conversion (series signal input and parallel signal output, i.e. SIPO).

\subsection{The logic scheme of trigger }\label{sec.III.2}

The trigger system conducts the following operations. 1) To construct and deliver the global trigger signal to each subsystem, 2) to switch different running mode of each subsystem upon users' requirement and 3) to provide the global control command and global standard time.  The trigger logic module is implemented in the FPGA of STM L1, STM L2 and MTM. 

%% Fig.TOF 2-level
\begin{figure*}[htb]
\centering
% \hspace{-0.7cm}
\includegraphics[width=0.85\textwidth]{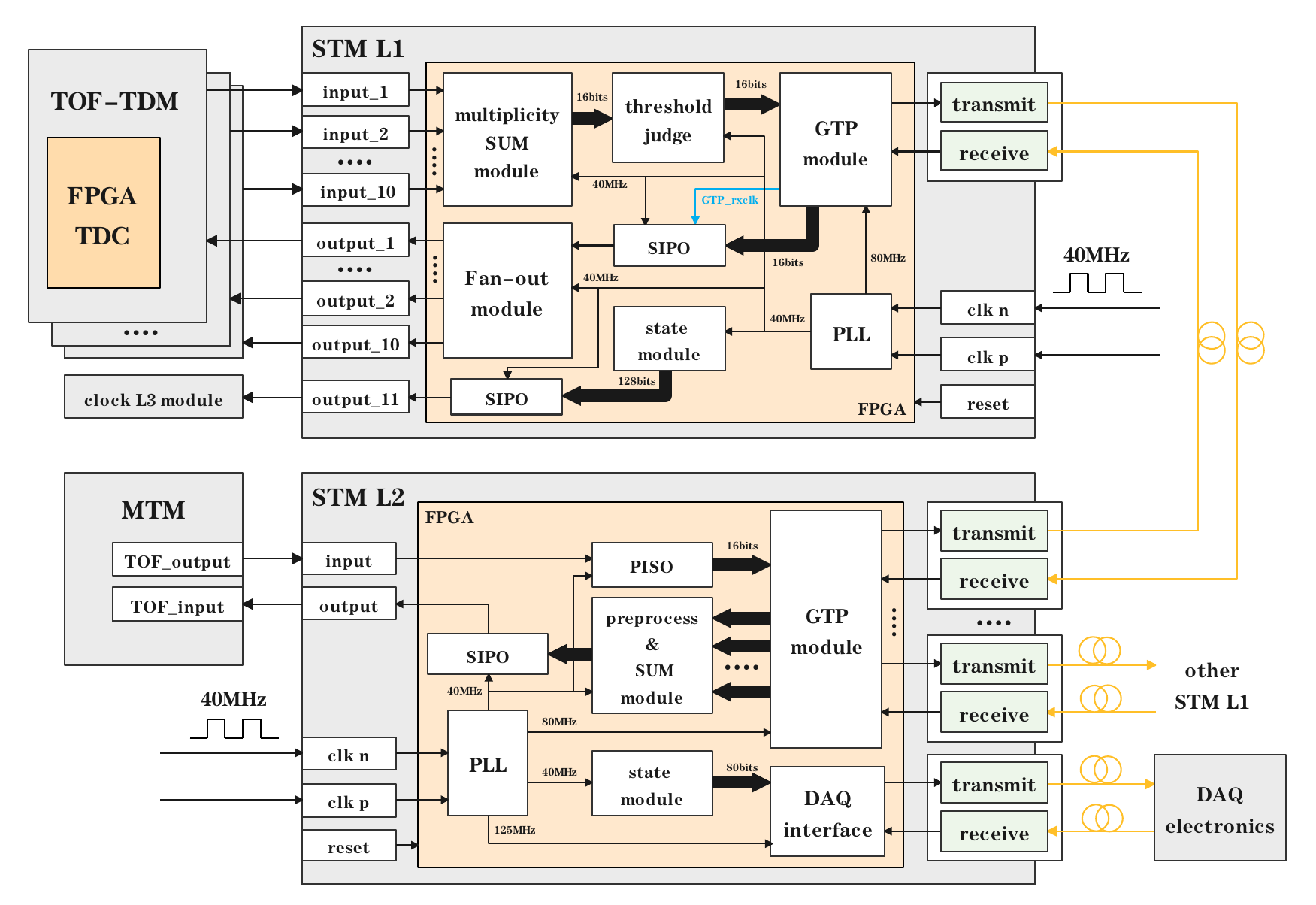}
\caption{(Color online) The 2-level trigger link STM L1 and STM L2 trigger logic and signal transmission diagram of iTOF and eTOF subsystem}
\label{TOF 2-level}
\end{figure*}

The logical details and signal flow of the 2-level trigger link of iTOF and eTOF subsystem are shown in FIG.\ref{TOF 2-level}. In the uplink chain, the local multiplicities in the TDMs are transformed to the STM L1 and  summed up in a 75ns time window in the "Multiplicity SUM module" of the STM L1 logic. In order to suppress the background noise, The summed multiplicity in each STM L1 is discriminated by  the "Threshold module"  before it is converted into an optical signal in GTP and sent to STM L2.  In STM L2, the multiplicity signals from all corresponding STMs L1 are summed  to obtain the total multiplicity of the physical event, as shown in the "preprocess \& SUM module" of STM L2 in FIG.\ref{TOF 2-level}. After the parallel-series conversion (parallel signal input and series signal output, PISO) is performed, it is uploaded to the MTM as an electrical signal. The format of the total multiplicity uplink data is 10 bits serial digital signal.

%% Fig.Tracking 2-level
\begin{figure*}[htb]
\centering
% \hspace{-0.7cm}
\includegraphics[width=0.85\textwidth]{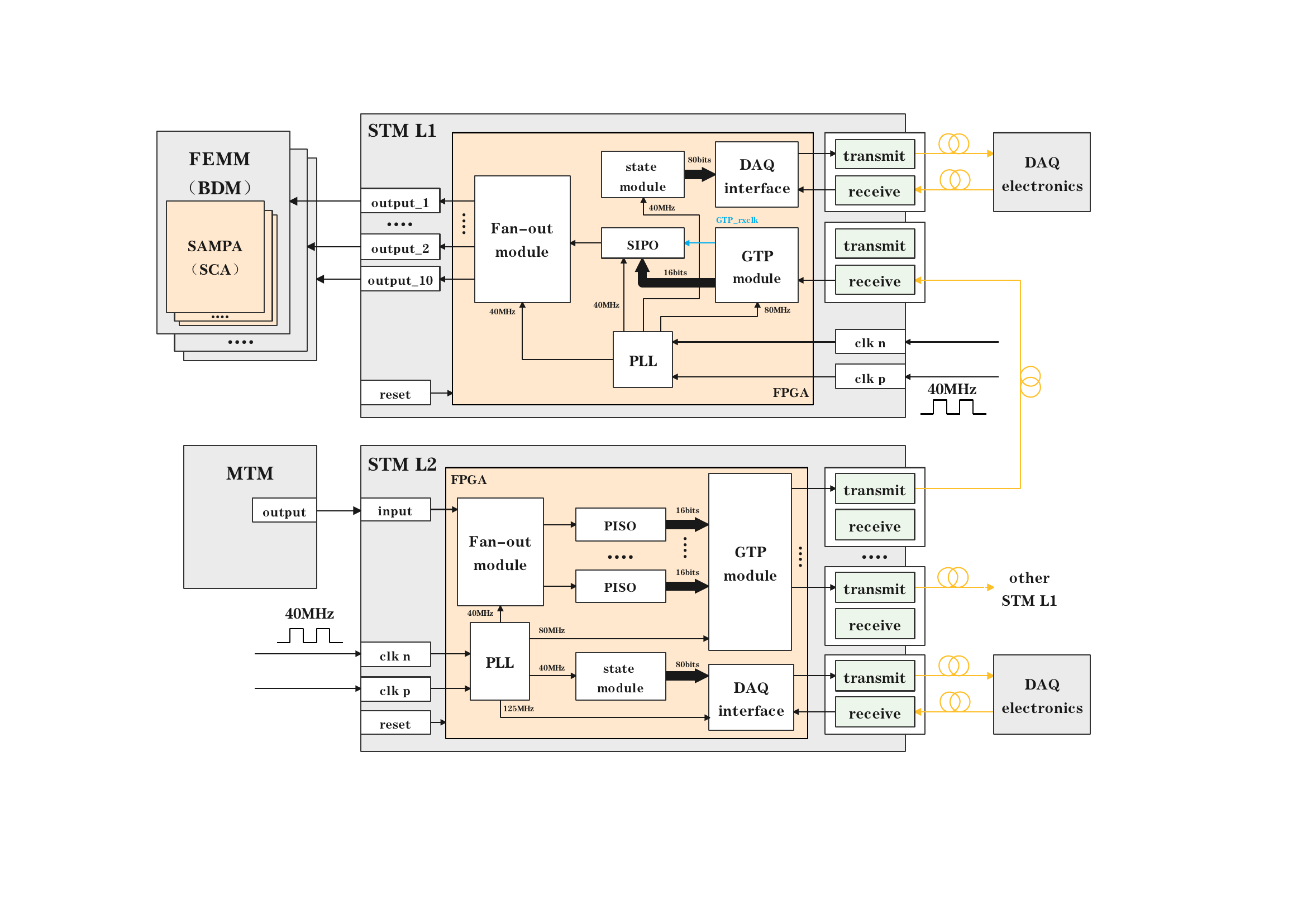}
\caption{(Color online) The 2-level trigger link STM L1 and STM L2 trigger logic and signal transmission diagram of TPC and MWDC subsystem}
\label{Tracking 2-level}
\end{figure*}

%% Fig.MTM
\begin{figure*}[htbp]
\centering
% \hspace{-0.7cm}
\includegraphics[width=0.95\textwidth]{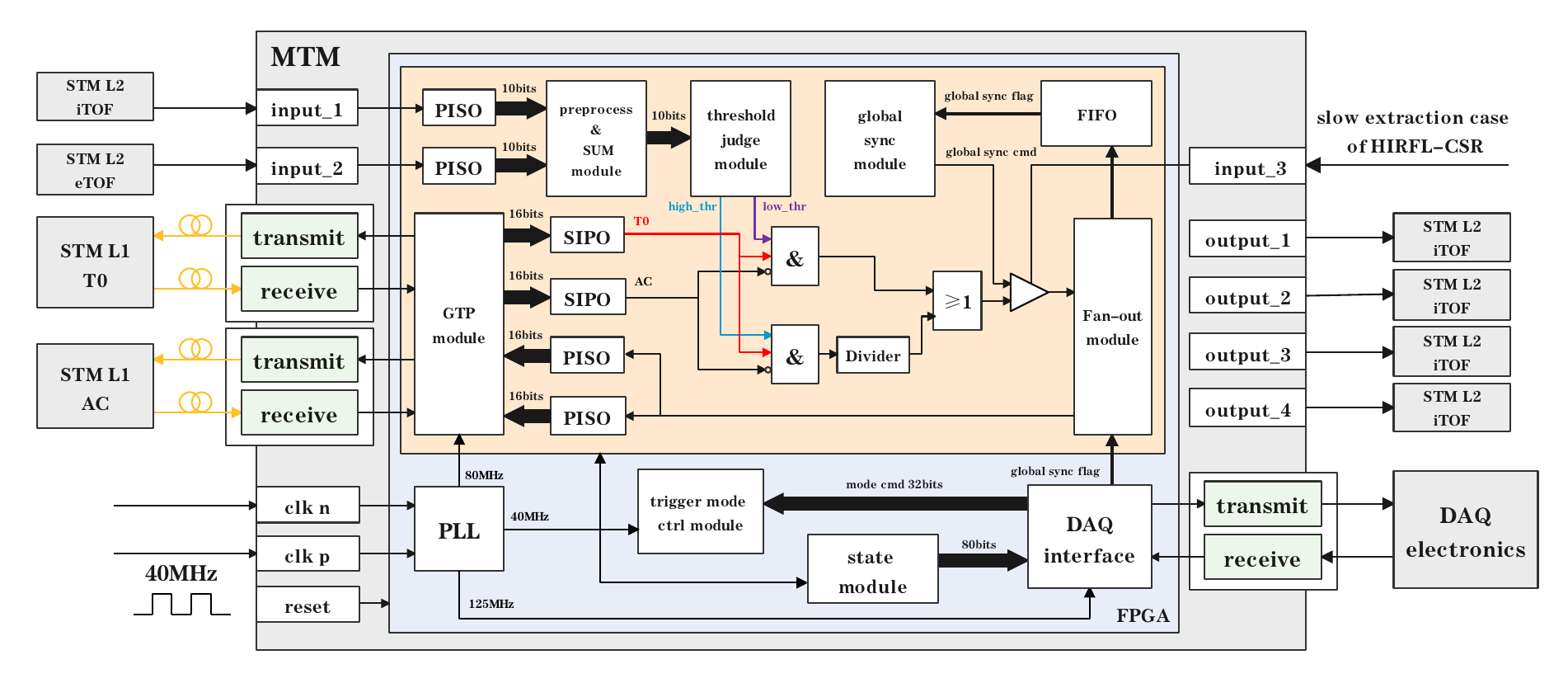}
\caption{(Color online) The trigger logic and signal transmission diagram of MTM}
\label{MTM}
\end{figure*}

Figure.\ref{Tracking 2-level} shows the logical details and the signal flow of the tracking system, using downlink chain only.  The global trigger signal generated from the MTM is sent to the STM L2 by electrical signal and fanned out in  multi channels, which are converted into an optical signal by GTP and distributed to each STM L1. In STM L1, the multi-channel signals are further fanned out and converted into serial signals sent to  the FEMMs. 

Subsystems involving  single level trigger links include ${\rm T_0}$, AC, ZDC, and beam monitor subsystems. Since  ${\rm T_0}$ and AC signals are included in the  calculation of global trigger, these two subsystems use both uplink and downlink transmission chain, while the ZDC and beam monitor subsystems need downlink chain only to receive trigger signal from MTM. 

Finally, the logical details and the signal flow  of MTM are shown in Figure.\ref{MTM}. The signal of ${\rm T_0}$ and AC are transmitted uplink from STM L1 to MTM via fiber. After the  multiplicity of iTOF and eTOF are transmitted to MTM, respectively, they are serialized and converted into 10-bit parallel signals. The total multiplicity of TOF is calculated in the "preprocess \& SUM module" and filtered  in the  "threshold judge module". The output of the threshold filtering is used to classify the event type and to generate the global trigger signal.
Due to the different time delay caused by the logical operations for TOF, ${\rm T_0}$ and AC, a further adjustable delay is enabled to synchronize all the subsystems. 

\subsection{Trigger mode and global control} \label{sec.III.2}

The trigger electronics are connected to the DAQ through optical fibers, and hence, the trigger system can be monitored and controlled  through the DAQ terminal.   The parameters used in the trigger logics, including the delays, coincidence gate width and the fraction divide factors, thresholds are all configurable remotely.   In addition, the function of remote logic update to the main trigger electronics is also implemented based on QuickBoot. The DAQ transfers and stores the new logic to the SPI Flash of MTM, and then sends the reconfiguration command to realize the remote logic update function. The SPI Flash model used by MTM is N25Q256A with a capacity of 256MB. DAQ can obtain the status information of the trigger electronics by reading the sensor measurements of each modules and crates, including temperature, humidity and current information.

The trigger system can change the running mode of CEE according to the system requirements through the DAQ control. In addition to the above beam mode, the cosmic ray mode and the electronic self-test mode are also included. In the cosmic ray mode  during beam-off, in which the ${\rm T_0}$ and AC do not participate in the trigger, MTM calculate the global trigger directly from the  the multiplicity information from iTOF or eTOF and  deliver to each subsystem. The electronic self-check mode is designed to check the status of the electronics of each system during the beam off in the frequency which is configurable by the user. In addition, the trigger system provides the hardware basis for the global control of the system. Namely, the control commands such as start acquisition, stop acquisition, and time synchronization  are sent to all systems via the same hardware links of the trigger system.

%% Fig.beam
\begin{figure*}[!htb]
\centering
% \hspace{-0.7cm}
\includegraphics[width=0.98\textwidth]{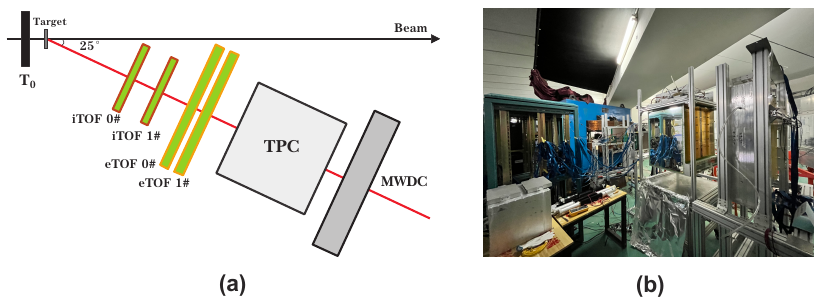}
\caption{(Color online) (a) Beam experiment detector position diagram. (b) Beam experiment site photos}
\label{beam}
\end{figure*}

\section{Beam test results} \label{sec. IV}

The trigger system has been tested with the prototypes of the ${\rm T_0}$, iTOF, eTOF, TPC and MWDC subsystems of CEE  in  beam experiment of 320MeV/u $^{56}$Fe+$^{56}$Fe at the second radioactive ion beam line at Lanzhou (RIBLL2). The setup of the beam test  is shown in FIG.\ref{beam}.  ${\rm T_0}$ is placed 20 cm in front of the target, and other detectors are placed in the direction with  $\theta_{lab}=25 ^{\circ}$ from the beam line. The high energy products of light charged particles pass through the iTOF, eTOF, TPC and MWDC, respectively. iTOF 0\# is placed 90 cm away from the target and 25cm away from iTOF 1\#. eTOF 0\# and iTOF 1\# are separated by 47 cm, the two eTOF are close together, and the TPC is placed  50 cm to the final TOF detector, and the TPC and MWDC are 50cm apart. The TPC readout plane consists of 3 layers of GEM foils and a pad plane.  

%% Fig.master chassis
\begin{figure}[htb]
\centering
% \hspace{-0.7cm}
\includegraphics[width=0.43\textwidth]{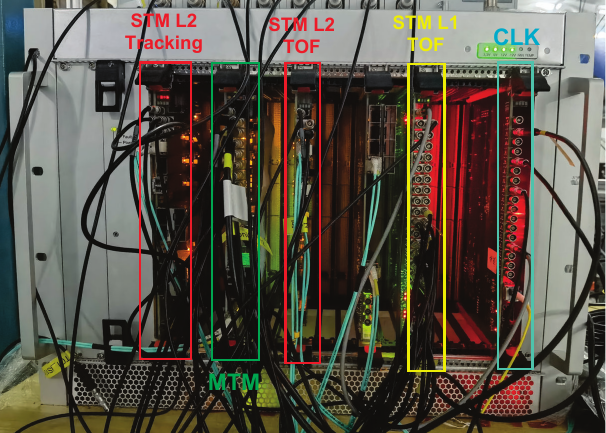}
\caption{(Color online) Beam experiment master chassis electronics setup}
\label{chassis}
\end{figure}

The 2-level trigger link of iTOF and eTOF subsystem, the 2-level link of TPC and MWDC subsystem, and the 1-level link of ${\rm T_0}$ subsystem are set up. Since this experiment is in the prototype test stage and the number of electronics channels is aboot 1k. The TOF subsystem is set to share one STM L1 and STM L2, the Tracking subsystem is set to share one STM L2, and the TPC and MWDC are set to have one STM L1 respectively in their FEMM chassis. FIG.\ref{chassis} shows the electronic settings of the main control box in the beam experiment site. The MTM and STM L2 are generally placed in the main control box, in addition to the DAQ data summary board and clock electronic. The DAQ data summary board connects the MTM to the PC and constructs the hardware link for remote control. The average beam count is about 10$^{5}$Hz  and the average trigger rate is about 1 kHz, while the peak trigger rate is about 4 kHz. The noise level of TOF detector is tested at 500Hz. The low and high noise threshold is set to  $M_{\rm l}=3$ and $M_{\rm h}=100$, respectively. The event classification threshold is set to $M_{\rm tof}=10$. The fraction division ratio is set to 1. 

%% Fig.wave
\begin{figure}[htb]
\centering
% \hspace{-0.7cm}
\includegraphics[width=0.44\textwidth]{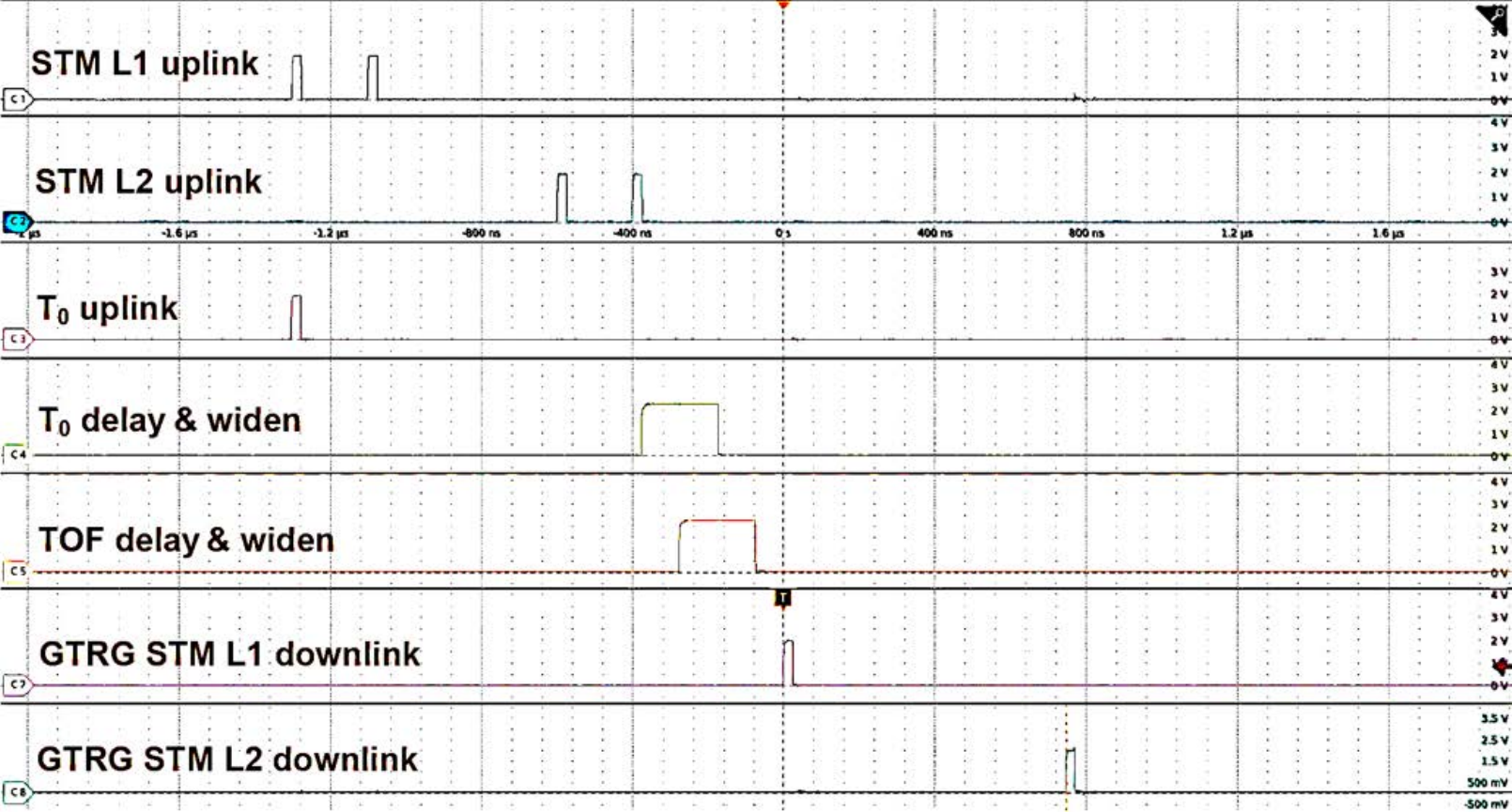}
\caption{(Color online) Uplink TOF multiplicity signal, ${\rm  T_0}$ signal and downlink global trigger signal waveform.}
\label{wave}
\end{figure}

In the experiment, the delay time of ${\rm  T_0}$ signal is set to 900 ns in MTM, the delay of TOF signal is unchanged, and the width of ${\rm  T_0}$ and TOF signal is extended to 200ns. FIG.\ref{wave} shows the timing relationship between the upstream signal involved in trigger calculations during the beam. The STM L1 uplink transmission serial signal of the TOF subsystem is 1'b1+$0000000100$, representing the signal with multiplicity of 4. The STM L2 of the TOF subsystem summarizes the multiplicity signal from STM L1 and displays the consistent output signal of STM L1. The delay of the 2-level signal is 700 ns. The channel 7 generates the global trigger signal for MTM, and channel 8 is the trigger signal for STM L1 downlink transmission. The delay time of the 2-level trigger link from MTM to STM L1 is 750 ns. Taking the iTOF subsystem as an example, the total delay of the trigger signal from TDM upstream transmission to STM L1 downstream transmission to TDM is $\rm 2.6 \mu s$.

\begin{figure}[htb]
\centering
% \hspace{-0.7cm}
\includegraphics[width=0.44\textwidth]{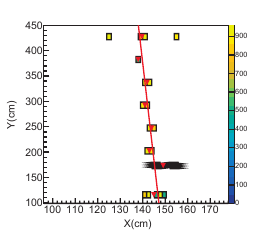}
\caption{(Color online) The particle tracks were obtained by fitting pad charge deposition at each layer of TPC}
\label{TPC_tracking}
\end{figure}

FIG.\ref{TPC_tracking} present an event example of a track detected in TPC in coincidence with the TOF detectors. The readout plane of TPC is designed in such way that different shapes of pad are applied to study the tracking resolution, including rectangular pad and ZigZag pad. A straight track is clearly seen. The red triangle symbols denote the gravity center of the signs on each corresponding row. The shadowed areas are the charged distribution induced on the pads.

%% Fig.TPC_tracking

%% Fig.time_res
\begin{figure}[htb]
\centering
% \hspace{-0.2cm}
\includegraphics[width=0.44\textwidth]{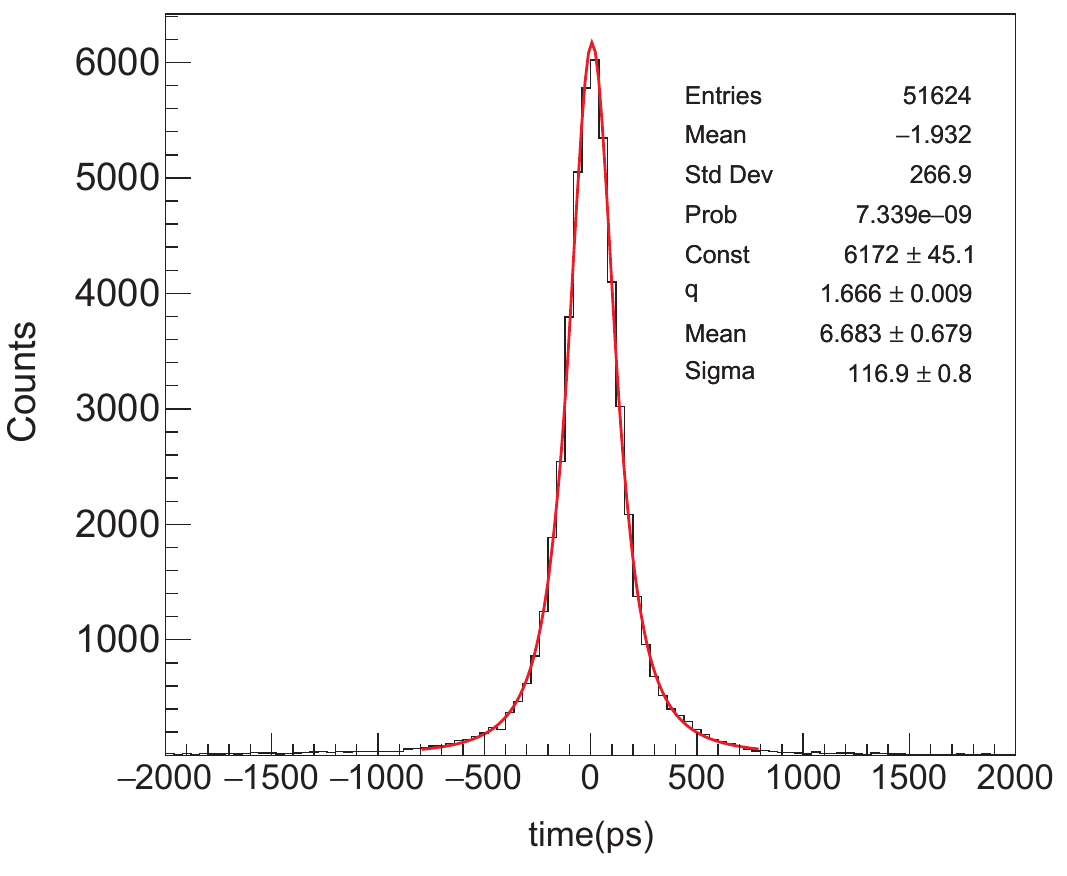}
\caption{(Color online) The TOF subsystem measured time-of-flight resolution.}
\label{time_res}
\end{figure}

The iTOF and eTOF subsystems provide the hits multiplicity information required in trigger signal construction. FIG.\ref{time_res} shows the distribution of the time difference between ${\rm T_0}$ and the fired TOF detector. The time resolution of on TOF prototype is about $117/\sqrt{2} \approx 83$ ps. Here the resolution  caused by the variation of the particles incident to the TOF detectors is not subtracted. The track reconstruction efficiency from the TOF hits is about $96\%$, slightly dependent on the firing detector. The results of the beam test demonstrate that the trigger system for CEE works correctly.

\section{Conclusion} \label{sec. V}

In this paper, the trigger system of HIRFL-CSR  external-target experiment (CEE)  is described. The trigger system has a master-slave structure, using optical fibers to transmit signals in the inter layer communication. The operations and calculations in the whole trigger scheme  at all levels are implemented  based on FPGA technology.  Different running modes, including beam experiment, cosmic ray calibration and electronics self-checks are implemented. Communications of the trigger system with DAQ and other system can be established. Results in the beam test, where the prototypes of various sub-detectors, DAQ and global clocks are all connected, demonstrate the trigger system functions correctly and meets the requirements of CEE.

\section*{Declaration of competing interest}
The authors declare that they have no known competing finan-cial interests or personal relationships that could have appeared to influence the work reported in this paper.

\section*{Acknowledgments}

 This work is supported  by the National Natural Science Foundation of China under Grant Nos. 11927901 and 11890712  %ZGX projects 
  and by Tsinghua University Initiative Scientific Research Program.

\bibliography{reference}

\end{document}